# Additional Boundary Condition for a Wire Medium Connected to a Metallic Surface

Mário G. Silveirinha[(1)], Carlos A. Fernandes[(2)], Jorge R. Costa[(2, 3)]

[(1)]Instituto de Telecomunicações-Universidade de Coimbra,
Departamento de Engenharia Electrotécnica, Pólo II, 3030 Coimbra, Portugal, mario.silveirinha@co.it.pt

[(2)]Instituto de Telecomunicações, Instituto Superior Técnico, Av. Rovisco Pais 1049-001 Lisboa, Portugal

[(3)] Instituto Superior de Ciências do Trabalho e da Empresa, Departamento de Ciências e Tecnologias da Informação, Av. das Forças Armadas, 1649-026 Lisboa, Portugal.

## Abstract

In this work, we demonstrate that the interaction of electromagnetic waves with a microstructured material formed by metallic wires connected to a metallic surface can be described using homogenization methods provided an additional boundary condition is considered. The additional boundary condition is derived by taking into account the specific microstructure of the wire medium. To illustrate the application of the result, we characterize a substrate formed by an array of tilted metallic wires connected to a ground plane, demonstrating that in such configuration the wire medium behaves essentially as a material with extreme optical anisotropy and that in some circumstances the substrate can be seen as an impedance surface.

# I. INTRODUCTION

In recent years, there has been a great interest in the design of novel materials with unusual properties such as negative permittivity and permeability [1], extreme optical anisotropy [2, 3], or permittivity near zero [4]. It has been demonstrated that novel microstructured materials (metamaterials) may have good potentials in subwavelength imaging applications [5], field concentration [6], and supermicroscopy [7, 8].

Unlike common materials, in some metamaterials the wavelength of operation is only 5-10 times larger than the characteristic size of the inclusions. Due to this reason, anomalous phenomena may occur in such materials. In particular, recent works have underlined the importance of spatial dispersion effects in several microstructured composites [9]-[13].

The electrodynamics of spatially dispersive materials is very peculiar. The nonlocal character of the material response may cause the emergence of "new" waves, as compared to the ordinary case in which only two plane waves propagate along each direction of space [14]. Thus, the problem of reflection and refraction at an interface becomes undetermined if one only uses the classical boundary conditions. To overcome this difficulty it may be necessary to introduce "additional boundary conditions" (ABC) [14]-[17]. The nature of the ABC depends on the specific microstructure of the material, and can be determined only on the basis of a microscopic model that describes the dynamics of the internal variables.

In a previous work [18], we derived an ABC that characterizes the scattering of electromagnetic waves by a wire medium. The wire medium consists of an array of long metallic parallel wires arranged in a periodic lattice. Previous studies demonstrated that this material is strongly spatially dispersive, even in the long wavelength limit [9, 10, 11]. Several key applications of wire media, such as the realization of materials with negative permittivity [1, 19] or the realization of near-field lenses with subwavelength resolution [5, 20], have been described in the literature. In [20], the ABC formulated in [18] was successfully applied to the study of the resolution of an imaging device formed by an array of nanorods operated at infrared frequencies. An alternative model to characterize the propagation of radiation at an interface of wire media has been suggested in [21], and is based on the introduction of a transition layer.

The ABC derived in [18] was obtained under the hypothesis that the material adjacent to the wire medium is non-conductive. However, in several configurations of interest the metallic wires may be connected to a ground plane. For example, textured and corrugated surfaces may have important applications in the design of high-impedance surfaces, suppression of guided modes, and realization of waveguides that support mode sizes below the diffraction limit [22]. When the wires are connected to a conductive material the ABC proposed in [18] does not apply.

The objective of this work is to generalize the results of [18], and derive a new ABC that characterizes the propagation of waves in a wire medium connected to a metallic ground plane. It is proven that due to the specific microstructure of the composite material, the macroscopic fields must be related in such a way that an additional boundary condition is verified. To validate the result, we study the reflection of plane waves by a substrate formed by an array of tilted metallic wires connected to a ground plane. It is demonstrated that the results of our homogenization model compare very well with full wave simulations. In addition, we study when the effects of spatial dispersion can be neglected. It is proved that in such circumstances the wire medium behaves as a material with extreme optical anisotropy, and that the microstructured substrate can be modeled as a surface impedance.

This paper is organized as follows. In section II, the required homogenization concepts are introduced and the new ABC is derived. In section III, the ABC is applied to characterize a substrate formed by an array of tilted wires connected to a metallic ground plane. In section IV, the proposed theory is tested against full wave simulations, and the response of the structured substrate is explained in terms of the extreme anisotropy of the wire medium. Finally, in section V the conclusions are drawn. It is assumed that the electromagnetic fields have a time dependence of the form $e^{j\omega t}$.

## II. THE ADDITIONAL BOUNDARY CONDITION

The wire medium considered here consists of an array of cylindrical metallic wires with radius $r_w$ arranged in a square lattice with lattice constant $a$. The wires are directed along the direction $\hat{\mathbf{u}}_\alpha = -\sin\alpha\,\hat{\mathbf{u}}_x + \cos\alpha\,\hat{\mathbf{u}}_z$, where $\hat{\mathbf{u}}_x$, $\hat{\mathbf{u}}_y$ and $\hat{\mathbf{u}}_z$ are the unit vectors along the coordinate axes. It is supposed that the wires are embedded in a dielectric host medium and connected (with a good Ohmic connection) to a perfect electric conducting (PEC) plane placed at $z=0$ (Fig. 1).

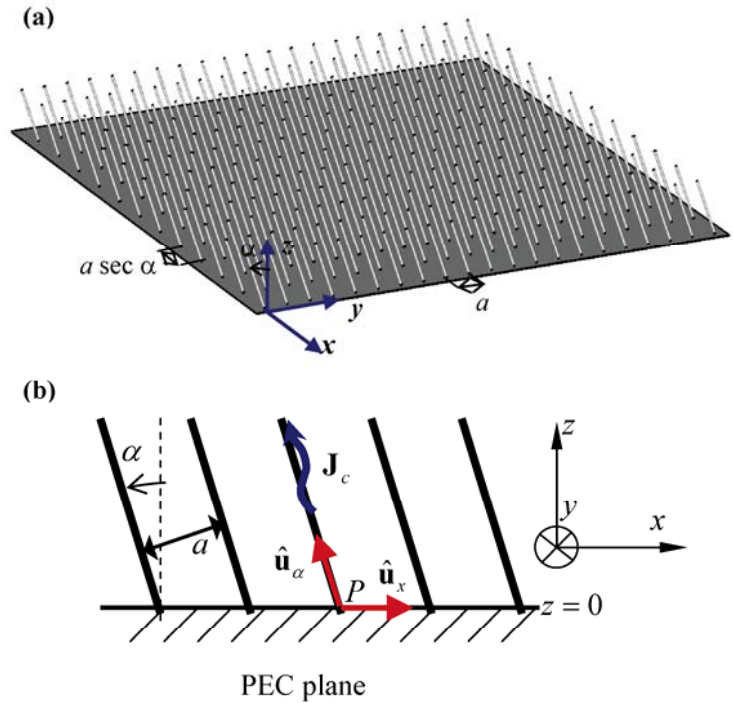

**Fig. 1** (Color online) (a) Tilted square array of wires connected to a ground plane. (b) Detail of the intersection of the tilted wire medium with the PEC ground plane.

The objective is to derive an additional boundary condition (ABC) for the macroscopic fields at the PEC interface. To this end, it is convenient to briefly introduce some homogenization concepts. Let $(\mathbf{e},\mathbf{b})$ denote the microscopic electric and induction fields in the wire medium slab. The density of current induced on the surface of a generic wire is given by $\mathbf{J}_c = \hat{\mathbf{v}} \times \mathbf{b}/\mu_0$, where $\hat{\mathbf{v}}$ is the unit vector normal to the pertinent wire. It is assumed that the fields have the Floquet property along the directions parallel to the PEC plane. More specifically, it is supposed that $(\mathbf{e},\mathbf{b})e^{j\mathbf{k}_\parallel \cdot \mathbf{r}}$ is a periodic function of $x$ and $y$, where $\mathbf{k}_\parallel = (k_x, k_y, 0)$ is the transverse wave vector, and $\mathbf{r} = (x,y,z)$ is a generic point of space. This situation occurs if for example the material is illuminated by a plane wave. As in [23], we define the transverse averaged (TA) macroscopic electric field as,

$$\mathbf{E}_{av,T}(z) = \frac{1}{A_{cell}} \int_{\Omega_T(z)} \mathbf{e}(\mathbf{r}) e^{j\mathbf{k}_\parallel \cdot \mathbf{r}} dx dy \tag{1}$$

where $\Omega_T$ represents the transverse unit cell in each $z$=const. plane, and $A_{cell} = a^2/\cos\alpha$ is the area of $\Omega_T$. The TA-induction field $\mathbf{B}_{av,T}$ is defined similarly. It was proven in [23] that the TA-fields verify the system,

$$\left(-j\mathbf{k}_\parallel + \frac{d}{dz}\hat{\mathbf{u}}_z\right) \times \mathbf{E}_{av,T} = -j\omega \mathbf{B}_{av,T} \tag{2a}$$

$$\left(-j\mathbf{k}_\parallel + \frac{d}{dz}\hat{\mathbf{u}}_z\right) \times \frac{\mathbf{B}_{av,T}}{\mu_0} = j\omega\varepsilon_0\varepsilon_h \mathbf{E}_{av,T}(z) + \mathbf{J}_{d,av}(z) \tag{2b}$$

where $\varepsilon_h$ is the relative permittivity of the host medium, and the average microscopic current for PEC inclusions is:

$$\mathbf{J}_{d,av}(z) = \frac{1}{A_{cell}} \int_{\partial A(z)} \mathbf{J}_c(\mathbf{r}) e^{j\mathbf{k}_\parallel \cdot \mathbf{r}} \frac{1}{|\hat{\mathbf{v}} \times \hat{\mathbf{u}}_z|} dl \tag{3}$$

In the above, $\partial A(z)$ is the contour defined by the intersection between the surface of the wire in the unit cell and the pertinent $z$=const. plane, and $dl$ is the element of arc.

In order to derive an ABC, we need to identify some property of the structure under study that can be used to obtain some non trivial relation between the macroscopic (average) fields.

For example, the ABC proposed in [18] was based on the fact that the microscopic current along the metallic wires must vanish at the interface between the wire medium slab and a non-conductive dielectric medium. However, as mentioned before, such property is not valid here because the metallic wires are connected to a PEC ground plane. However, let us analyze the properties of the electric density of surface charge $\sigma_c$ near the connection (point $P$) between a generic wire and the PEC plane (panel (b) of Fig. 1).

Since the tangential electric field vanishes at a PEC surface, it is clear that at the junction point $P$, we have that $\mathbf{e}.\hat{\mathbf{u}}_\alpha = 0$ (since point $P$ belongs to the metallic wire), and $\mathbf{e}.\hat{\mathbf{u}}_x = 0$, $\mathbf{e}.\hat{\mathbf{u}}_y = 0$ (since point $P$ belongs to the ground plane). Thus, the electric field must vanish at $P$, i.e. $\mathbf{e} = \mathbf{0}$. On the other hand, the density of surface charge on a point of the wire is given by $\sigma_c = \varepsilon_0 \varepsilon_h \hat{\mathbf{v}}.\mathbf{e}$. Hence, it is evident that the density of electric charge $\sigma_c$ in a generic wire must vanish at the connection with the PEC ground plane. This is the key property that will be used to derive a new ABC for an interface between a wire medium and a PEC ground plane.

To this end, it is noted that $\sigma_c = -\nabla_s.\mathbf{J}_c / j\omega$, where "$\nabla_s.$" represents the surface divergence of the vector field. For thin wires, it is a good approximation to consider that the density of current flows along the direction of the wire axis (parallel to $\hat{\mathbf{u}}_\alpha$). Within this hypothesis, we have that $\sigma_c = -\dfrac{1}{j\omega}\dfrac{dJ_{c,\alpha}}{ds}$, where $J_{c,\alpha} = \mathbf{J}_c.\hat{\mathbf{u}}_\alpha$ and $s$ is the wire length measured along $\hat{\mathbf{u}}_\alpha$ ($s = 0$ at point $P$). The previous results demonstrate that in order that $\sigma_c = 0$ at the connection points it is necessary that $\dfrac{dJ_{c,\alpha}}{ds} = 0$.

To proceed we need to relate the microscopic current $\mathbf{J}_c$ with the macroscopic average fields. Within the thin wire approximation we may assume that the density of surface current in the metallic wires is of the form, $\mathbf{J}_c = \dfrac{I(z)}{2\pi r_w} e^{-j\mathbf{k}_\parallel \cdot \mathbf{r}} \hat{\mathbf{u}}_\alpha$, where $I$ is the electric current along the wire. Hence, simple calculations show that $\mathbf{J}_{d,\text{av}}$ given by (3) is equal to,

$$\mathbf{J}_{d,\text{av}}(z) = \dfrac{1}{a^2} I(z) \hat{\mathbf{u}}_\alpha \qquad (4)$$

This result implies that,

$$J_{c,\alpha} = \dfrac{a^2}{2\pi r_w} e^{-j\mathbf{k}_\parallel \cdot \mathbf{r}} \mathbf{J}_{d,\text{av}}(z) \cdot \hat{\mathbf{u}}_\alpha \qquad (5)$$

Using the fact that $\mathbf{r} = s\hat{\mathbf{u}}_\alpha$ along the axis of the considered wire, it is seen that in order that $\dfrac{dJ_{c,\alpha}}{ds} = 0$ at the connection points it is necessary the following *additional boundary condition* is verified at the PEC plane:

$$\left(-j\mathbf{k}_\parallel + \hat{\mathbf{u}}_z \dfrac{d}{dz}\right) \cdot \hat{\mathbf{u}}_\alpha \hat{\mathbf{u}}_\alpha \cdot \mathbf{J}_{d,\text{av}} = 0 \qquad (6)$$

It is stressed that this property is a mere consequence of the density of charge $\sigma_c$ being zero at the extremity of the wires connected to the ground plane, and of the thin wire approximation. Using (2b), the ABC can be rewritten in terms of the macroscopic fields. It is readily found that:

$$\left(\mathbf{k}_\parallel \cdot \hat{\mathbf{u}}_\alpha + \hat{\mathbf{u}}_z \cdot \hat{\mathbf{u}}_\alpha j \dfrac{d}{dz}\right)\left(\omega \varepsilon_0 \varepsilon_h \hat{\mathbf{u}}_\alpha \cdot \mathbf{E}_{\text{av,T}} + \hat{\mathbf{u}}_\alpha \times \left(\mathbf{k}_\parallel + \hat{\mathbf{u}}_z j \dfrac{d}{dz}\right) \cdot \dfrac{\mathbf{B}_{\text{av,T}}}{\mu_0}\right) = 0 \quad \text{(at PEC plane)} \qquad (7)$$

In the next section, it will be confirmed that this boundary condition together with the classical boundary condition, $\hat{\mathbf{u}}_z \times \mathbf{E}_{\text{av,T}} = 0$, completely characterize the reflection of waves by a wire medium slab adjacent to a conducting plane.

As mentioned before and proved in [18], an ABC is also required at an interface between the wire medium and a dielectric material (e.g. air). The ABC obtained in [18] was derived under the assumption that the wires are normal to the interface. In the rest of this section, we will prove that such ABC is also valid even when the wires are tilted with respect to the interface. As discussed in [18], at an interface with a non-conductive material and for thin wires it is necessary that $J_{c,\alpha} = \mathbf{J}_c . \hat{\mathbf{u}}_\alpha = 0$. Thus, from (5) the ABC at an interface with air is:

$$\mathbf{J}_{d,\text{av}} . \hat{\mathbf{u}}_\alpha = 0 \qquad \text{(interface with air)} \qquad (8)$$

As proved next, the above condition is equivalent to the ABC derived in [18]:

$$\varepsilon_h \mathbf{E}_{\text{av,T}} . \hat{\mathbf{u}}_z \Big|_{\text{wire medium side}} = \mathbf{E}_{\text{av,T}} . \hat{\mathbf{u}}_z \Big|_{\text{air side}} \qquad (9)$$

To obtain this result, first we note that if $\hat{\mathbf{u}}_\perp$ is an arbitrary vector normal to $\hat{\mathbf{u}}_\alpha$, i.e. $\hat{\mathbf{u}}_\perp . \hat{\mathbf{u}}_\alpha = 0$, then from (4) it is clear that $\mathbf{J}_{d,\text{av}}(z) . \hat{\mathbf{u}}_\perp = 0$. It is crucial to notice that the condition $\mathbf{J}_{d,\text{av}}(z) . \hat{\mathbf{u}}_\perp = 0$ is not an ABC. In fact, it is valid at every *z=const.* plane inside the wire medium, not only at the interface. It is a trivial consequence of the induced current inside the material flowing solely along the direction $\hat{\mathbf{u}}_\alpha$, and is satisfied automatically by every solution of the Maxwell's equations. It is clear that the relation $\mathbf{J}_{d,\text{av}}(z) . \hat{\mathbf{u}}_\perp = 0$ and equation (8), imply that for $\alpha \neq 90^\circ$,

$$\mathbf{J}_{d,\text{av}} . \hat{\mathbf{u}}_z = 0 \qquad \text{(interface with air)} \qquad (10)$$

To proceed, it is noted that the above condition is valid both at the air side and at the wire medium side of the interface (this is obvious because $\mathbf{J}_{d,\text{av}}$ vanishes in the air region). Thus, equation (10) is equivalent to $[\mathbf{J}_{d,\text{av}}] . \hat{\mathbf{u}}_z = 0$, where $[...]$ represents the jump discontinuity of the quantity inside brackets at the interface. But using (2b) and the fact that the tangential induction field is continuous at the interface [23], it is clear that $[\mathbf{J}_{d,\text{av}}] . \hat{\mathbf{u}}_z = -j\omega\varepsilon_0 [\varepsilon_h(z) \mathbf{E}_{\text{av,T}}(z)] . \hat{\mathbf{u}}_z$, where by definition $\varepsilon_h(z) = 1$ at the air side. Thus, since $[\mathbf{J}_{d,\text{av}}] . \hat{\mathbf{u}}_z = 0$ it follows that equation (9) is verified by the macroscopic fields, as we wanted to prove. To conclude, we note that formula (9) is not equivalent to the continuity of the electric displacement vector, since the effective permittivity of the wire medium is not $\varepsilon_h$.

# III. SCATTERING PROBLEM

In order to illustrate the application of the new ABC here we study the properties of a microstructured substrate formed by tilted metallic pins connected to a PEC plane. The wires are embedded in a dielectric substrate with relative permittivity $\varepsilon_h$ and height $T$, as illustrated in Fig. 2. It is also assumed that the metallic pins lie in planes parallel to the $xoz$ plane, separated by the distance $a$. The angle between the wires and the $z$-direction is $\alpha$. Such configuration is a generalization of that considered in [24, 25] for the case of vertical wires.

The objective is to obtain the reflection characteristic of the structure for TM-plane wave incidence (incident magnetic field is parallel to the PEC ground plane). For simplicity, we restrict our study to the case in which the metallic wires are parallel to the plane of incidence formed by the normal to the surface ($\hat{\mathbf{u}}_z$) and $\mathbf{k}^{inc}$ (see Fig. 2).

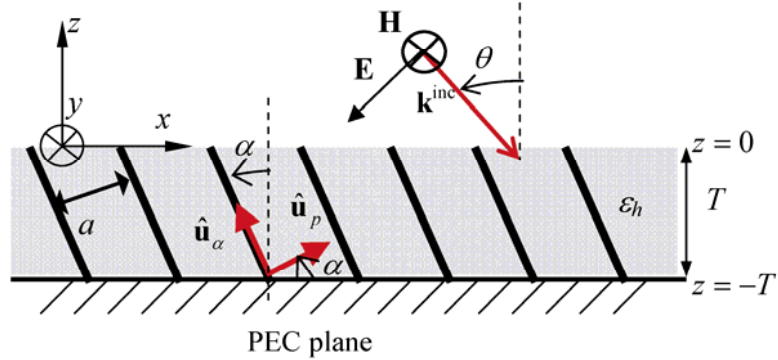

**Fig. 2** (Color online) Geometry of the wire medium substrate formed by a tilted square array of wires. The metallic wires lie in planes parallel to the $xoz$ plane. The spacing between the wires is $a$.

The wire medium is characterized by the dielectric function [9, 10],

$$\bar{\bar{\varepsilon}}_{eff} = \varepsilon_0 \varepsilon_h \left( \hat{\mathbf{u}}_p \hat{\mathbf{u}}_p + \hat{\mathbf{u}}_y \hat{\mathbf{u}}_y + \varepsilon_{\alpha\alpha}(\omega, k_\alpha) \hat{\mathbf{u}}_\alpha \hat{\mathbf{u}}_\alpha \right), \qquad \varepsilon_{\alpha\alpha}(\omega, k_\alpha) = 1 - \frac{\beta_p^2}{\varepsilon_h \omega^2/c^2 - k_\alpha^2} \tag{11}$$

where the unit vectors $\hat{\mathbf{u}}_\alpha = -\sin\alpha\,\hat{\mathbf{u}}_x + \cos\alpha\,\hat{\mathbf{u}}_z$ and $\hat{\mathbf{u}}_p = \cos\alpha\,\hat{\mathbf{u}}_x + \sin\alpha\,\hat{\mathbf{u}}_z$ are represented in Fig. 2, $c$ is the speed of light in vacuum, $k_\alpha = \mathbf{k}.\hat{\mathbf{u}}_\alpha$ is the $\alpha$-component of the wave vector

$\mathbf{k} = (k_x, k_y, k_z)$ of a plane wave, and $\beta_p$ is the plasma wave number, which only the depends on the geometrical properties of the lattice:

$$(\beta_p a)^2 = \frac{2\pi}{\ln\left(\frac{a}{2\pi r_w}\right) + 0.5275} \quad (12)$$

This homogenization model was derived in [9, 10] for a crystal formed by infinitely long parallel metallic wires (bulk medium). In particular, the macroscopic electromagnetic fields $(\mathbf{E}_{av}, \mathbf{B}_{av})$ associated with this model are averaged over the unit cell, which is a volumetric region [10]. Thus, it is not obvious that the homogenization model can be used to characterize the transverse average fields (1) introduced in the previous section. Notice that unlike the bulk medium fields, the TA-fields are averaged over a cross-section of the unit cell which is a surface [23]. In the Appendix it is demonstrated formally, that the TA-fields can be identified with the bulk medium fields for this particular material. Hence, in the rest of this work we will drop the subscripts "av" and "av,T", and we will denote the macroscopic fields simply by $\mathbf{E}$ and $\mathbf{H}$, where $\mathbf{H}$ is the macroscopic induction field divided by $\mu_0$.

The incoming wave is characterized by the normalized magnetic field $\mathbf{H}^{inc} = e^{+\gamma_0 z} e^{-jk_x x} \hat{\mathbf{u}}_y$, where $k_x$ is the component of the incident wave vector parallel to the interface, and $\gamma_0 = \sqrt{k_x^2 - (\omega/c)^2}$. For a propagating wave, $k_x$ is such that $k_x = (\omega/c)\sin\theta$, where $\theta$ is the angle of incidence.

It can be verified that the considered hypotheses imply that the magnetic field only has a y-component in all space: $\mathbf{H} = H_y \hat{\mathbf{u}}_y$. Moreover, inside the material slab the magnetic field is written in terms of the TM and TEM plane-wave like modes supported by the wire medium [9, 10]. The dispersion characteristic of these modes is,

$$\frac{\omega}{c}\sqrt{\varepsilon_h} = \pm \mathbf{k} \cdot \hat{\mathbf{u}}_\alpha \quad \text{(TEM mode)} \quad (13a)$$

$$\varepsilon_h \left(\frac{\omega}{c}\right)^2 = \beta_p^2 + k^2 \quad \text{(TM mode)} \tag{13b}$$

where $k^2 = \mathbf{k}.\mathbf{k}$, and $\mathbf{k} = (k_x, k_y, k_z)$ is the wave vector. It is well known that the transverse component $k_x$ of the incident wave vector is preserved. Thus, the modes excited inside the wire medium must be associated with a wave vector of the form $\mathbf{k} = \left(k_x, 0, k_z^{(i)}\right)$, where $k_x$ only depends on the incident wave, and $k_z^{(i)}$ depends on the considered mode and is obtained from the dispersion characteristic (13). For the case of a TM-wave it is straightforward to verify that $k_z^{(i)} = \pm j\gamma_{TM}$, where $\gamma_{TM} = \sqrt{\beta_p^2 + k_x^2 - \varepsilon_h (\omega/c)^2}$ is the attenuation constant along the z-direction. We note that $\gamma_{TM}$ is independent of the $\alpha$-angle that defines the orientation of the metallic wires. On the other hand, from (13a) it is found that for a TEM wave $k_z^{(i)} = k_{z,TEM}^+$ or $k_z^{(i)} = k_{z,TEM}^-$, where $k_{z,TEM}^+$ and $k_{z,TEM}^-$ are defined by:

$$k_{z,TEM}^\pm = \frac{\pm\sqrt{\varepsilon_h}\,\omega/c + k_x \sin\alpha}{\cos\alpha} \tag{14}$$

In the particular case $\alpha = 0$, the two propagation constants are symmetric, and the phase velocity along z is the same as the speed of light in the substrate material. However, when $\alpha \neq 0$ the propagation constants $k_{z,TEM}^+$ and $k_{z,TEM}^-$ are not symmetric, and thus the phase velocity along the positive z-direction is different from the phase velocity along the negative z-direction. This is a consequence of the anisotropy and orientation of the optical axes of the wire medium.

From the previous considerations, it follows that the magnetic field in all space is given by:

$$H_y = \left(e^{+\gamma_0 z} + \rho\, e^{-\gamma_0 z}\right) e^{-jk_x x}, \qquad \text{air side: } z > 0 \tag{15a}$$

$$H_y = \left(B_{TEM}^+ e^{-jk_{z,TEM}^+ z} + B_{TEM}^- e^{-jk_{z,TEM}^- z} + B_{TM}^+ e^{-\gamma_{TM} z} + B_{TM}^- e^{+\gamma_{TM} z}\right) e^{-jk_x x},$$
$$\text{(wire medium slab: } -T < z < 0) \tag{15b}$$

where $\rho$ is the reflection coefficient, and $B_{TEM}^{\pm}$ and $B_{TM}^{\pm}$ are the amplitudes of the excited TEM and TM waves, respectively. To obtain the electric field, we note that for a TM or TEM plane wave mode with magnetic field $\mathbf{H}(\mathbf{r};\mathbf{k}) = H_{y0} e^{-j\mathbf{k}\cdot\mathbf{r}} \hat{\mathbf{u}}_y$, the corresponding electric field is given by,

$$\mathbf{E}(\mathbf{r};\mathbf{k}) = \frac{1}{j\omega} \overline{\overline{\varepsilon}}_{eff}^{-1} \cdot \nabla \times \mathbf{H} =$$

$$= \frac{1}{j\omega} \overline{\overline{\varepsilon}}_{eff}^{-1} \cdot \left(-j\mathbf{k} \times \hat{\mathbf{u}}_y\right) H_{y0} e^{-j\mathbf{k}\cdot\mathbf{r}} \quad (16)$$

$$= \frac{1}{j\omega\varepsilon_0 \varepsilon_h} \left(\hat{\mathbf{u}}_p \hat{\mathbf{u}}_p + \frac{1}{\varepsilon_{\alpha\alpha}(\omega, k_\alpha)} \hat{\mathbf{u}}_\alpha \hat{\mathbf{u}}_\alpha\right) \cdot \left(-j\mathbf{k} \times \hat{\mathbf{u}}_y\right) H_{y0} e^{-j\mathbf{k}\cdot\mathbf{r}}$$

where $\varepsilon_{\alpha\alpha}(\omega, k_\alpha)$ is defined as in (11). Thus, after some calculations it is found that the electric field inside the wire medium slab is,

$$\mathbf{E} = \frac{\eta_0}{\sqrt{\varepsilon_h}} \left(B_{TEM}^+ e^{-jk_{z,TEM}^+ z} - B_{TEM}^- e^{-jk_{z,TEM}^- z}\right) e^{-jk_x x} \hat{\mathbf{u}}_p$$

$$+ \frac{\eta_0 c}{\varepsilon_h \omega} \left[\left(-j\gamma_{TM} + \frac{\beta_p^2}{k_p^+} \sin\alpha\right) \hat{\mathbf{u}}_x - \left(k_x + \frac{\beta_p^2}{k_p^+} \cos\alpha\right) \hat{\mathbf{u}}_z\right] B_{TM}^+ e^{-\gamma_{TM} z} e^{-jk_x x} \quad (17a)$$

$$+ \frac{\eta_0 c}{\varepsilon_h \omega} \left[\left(+j\gamma_{TM} + \frac{\beta_p^2}{k_p^-} \sin\alpha\right) \hat{\mathbf{u}}_x - \left(k_x + \frac{\beta_p^2}{k_p^-} \cos\alpha\right) \hat{\mathbf{u}}_z\right] B_{TM}^- e^{+\gamma_{TM} z} e^{-jk_x x}$$

where $k_p^{\pm} = k_x \cos\alpha - (\pm j\gamma_{TM}) \sin\alpha$, and $\eta_0 = \sqrt{\mu_0/\varepsilon_0}$ is the free-space impedance. Similarly, it can be verified that the electric field in the air region is:

$$\mathbf{E} = \eta_0 \frac{c}{\omega} \left(\left(+j\gamma_0 \hat{\mathbf{u}}_x - k_x \hat{\mathbf{u}}_z\right) e^{+\gamma_0 z} + \rho \left(-j\gamma_0 \hat{\mathbf{u}}_x - k_x \hat{\mathbf{u}}_z\right) e^{-\gamma_0 z}\right) e^{-jk_x x} \quad (17b)$$

To calculate the unknown coefficients, $\rho$, $B_{TEM}^{\pm}$, and $B_{TM}^{\pm}$, we need to impose boundary conditions at the interfaces $z = 0$ and $z = -T$. The classic boundary conditions establish that the tangential electric and magnetic fields, $E_x$ and $H_y$, are continuous at $z = 0$ (interface with air), and that $E_x$ vanishes at $z = -T$ (interface with PEC plane). These boundary conditions are clearly insufficient to calculate the unknown coefficients, because there are 5 unknowns, ($B_{TEM}^{\pm}$ and $B_{TM}^{\pm}$ and $\rho$), and the classical boundary conditions only yield 3 independent equations. It is thus evident the need of the ABCs derived in section II.

At the interface with air, it is thus necessary to impose the ABC (9), in addition to the continuity of the tangential fields:

$$[E_x] = 0, \qquad [H_y] = 0, \qquad [\varepsilon_h(z) E_z] = 0, \quad \text{at } z = 0 \qquad (18)$$

As before, $[...]$ represents the jump discontinuity of the function inside rectangular brackets at the interface, and $\varepsilon_h(z) = 1$ at the air side and $\varepsilon_h(z) = \varepsilon_h$ at the wire medium side. On the other hand, at the PEC interface the macroscopic fields must satisfy the new ABC (7). For the geometry under study, this ABC is equivalent to:

$$\left( j\cos\alpha \frac{d}{dz} - k_x \sin\alpha \right)\left( -\frac{\omega}{c}\varepsilon_h \sin\alpha \frac{E_x}{\eta_0} + \frac{\omega}{c}\varepsilon_h \cos\alpha \frac{E_z}{\eta_0} + \left( k_x \cos\alpha + j\sin\alpha \frac{d}{dz} \right) H_y \right) = 0$$
$$\text{at } z = -T \qquad (19a)$$

In addition, it is imposed that the tangential electric field vanishes at the ground plane:

$$E_x = 0, \qquad\qquad z = -T \qquad (19b)$$

Using the field distributions (15), (17) and the boundary conditions (18) and (19), we can easily obtain a determined linear system for the 5 unknowns of the problem. This system can be solved either numerically or analytically. These results will be described in the next section. It is interesting to mention that when the wires are normal to the ground plane and $\alpha = 0$, the boundary conditions (19) are equivalent to impose that the electric fields associated with the TM and TEM modes vanish independently at the PEC plane [25]. Such result is demonstrated formally in Appendix B.

## IV. DISCUSSION

In order to validate the new ABC and the proposed homogenization model, we numerically calculated the reflection coefficient as a function of frequency using the full wave electromagnetic simulator CST Microwave Studio[TM]. In the simulations, it was assumed that the radius of the wires is $r_w = 0.05a$. The results of the full wave simulations for different configurations of the structured substrate are depicted in Fig. 3 and Fig. 4 (discrete symbols) superimposed on the results obtained using the analytical model (solid lines) described in the previous section. It is seen that the overall agreement is excellent, even for wide incident angles, a relatively high $\varepsilon_h$, for tilt angles as large as $\alpha = 60°$ (Fig. 4), and for both short and long wires. The agreement deteriorates for $\omega a / c > 1.5$ since the long wavelength approximation becomes less accurate. Similar results are obtained for other configurations. Notice that the length of the metallic pins, $L_w$, is related to the thickness of the substrate, $T$, through the formula $L_w = T \sec \alpha$ where "sec" is the trigonometric secant function.

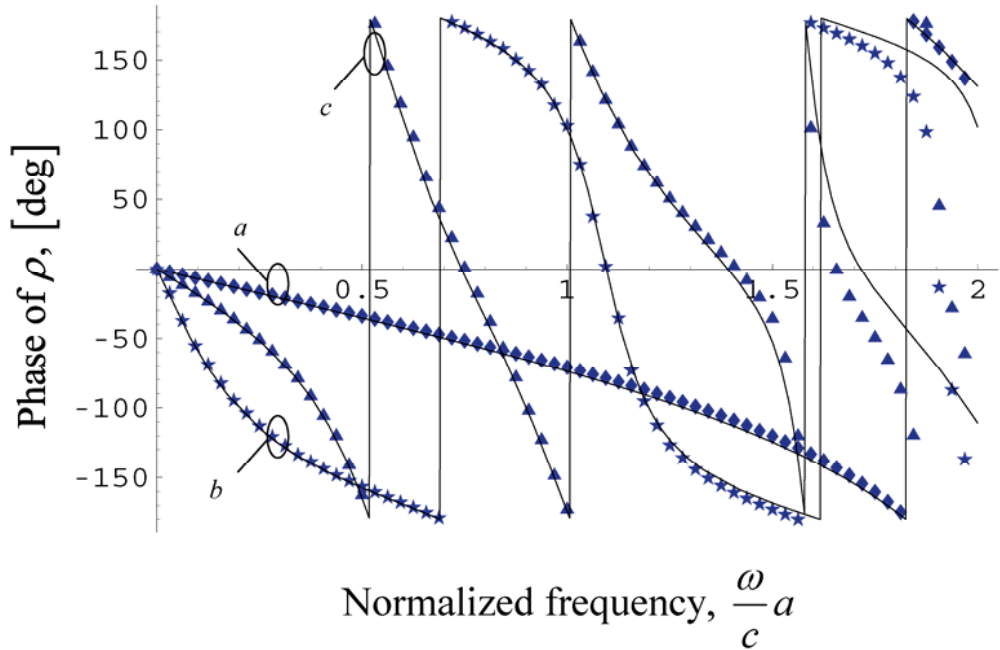

**Fig. 3** (Color online) Phase of the reflection coefficient as a function of the normalized frequency. Solid lines: our analytical model. Discrete symbols: numerical results obtained with a full wave simulator. The radius of the wires is $r_w = 0.05a$ and the tilt angle is $\alpha = 45°$. The parameters of the problem are: (*a*) permittivity of the dielectric substrate: $\varepsilon_h = 1.0$; thickness of the substrate: $T = 0.65a$; angle of incidence: $\theta = 45°$, (*b*) $\varepsilon_h = 2.2$; $T = 1.2a$; $\theta = 80°$, (*c*) $\varepsilon_h = 4.0$; $T = 1.2a$; $\theta = 45°$.

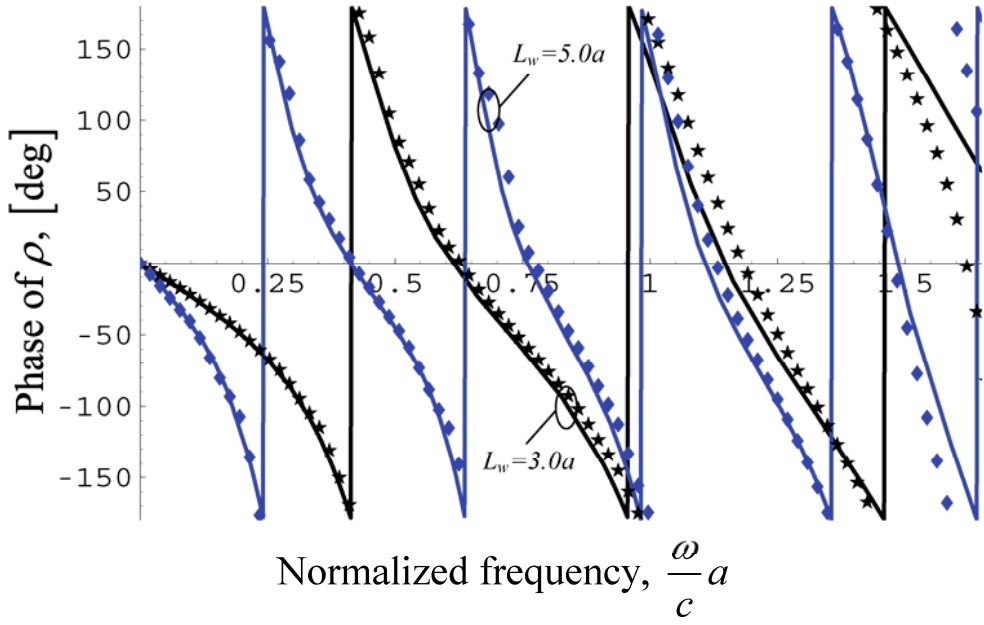

**Fig. 4** (Color online) Similar to Fig. 3, but for the parameters $\alpha = 60º$, $\theta = 45º$, $r_w = 0.05a$, $\varepsilon_h = 2.2$, and different values of $L_w = T \sec \alpha$.

The physical response of the array of tilted wires can be better understood by noting that the array of microstructured wires behaves approximately as a material with extreme anisotropy. Indeed, it is known that the TEM mode supported by the wire medium sees an effective permittivity of the form [9],

$$\overline{\overline{\varepsilon}}_{eff} = \varepsilon_0 \varepsilon_h \left( \hat{\mathbf{u}}_p \hat{\mathbf{u}}_p + \hat{\mathbf{u}}_y \hat{\mathbf{u}}_y + \infty \hat{\mathbf{u}}_\alpha \hat{\mathbf{u}}_\alpha \right) \tag{20}$$

i.e. the permittivity seen by the TEM mode along the direction of the wire axes is infinite. Materials with extreme optical anisotropy have remarkable properties [3], and in particular may be operated in a canalization regime in which a given field distribution is transferred from a source plane to an image plane with subwavelength resolution [5]. In practice, due to spatial dispersion effects, the permittivity model (20) is only a rough approximation of the actual nonlocal dielectric function (11) of the material. The spatial dispersion effects can be ignored only if the effect of the TM mode is of second order. Since the attenuation constant of the TM mode $\gamma_{TM}$ is roughly proportional to $1/a$ ($a$ is the spacing between the wires), it is clear that the spatial dispersion effects are negligible only in the limit $a/L_w \to 0$, where $L_w$ is the length of the wires. Or in other words, the spatial dispersion effects are negligible only if the wires are very densely packed. In this limit the material can be accurately described by the permittivity model (20). It is interesting to note that wire media can also be used to emulate media with negative permittivity [1, 9, 11, 19]. For such application the wires are oriented in such a way that the TEM mode cannot be excited and operated at a frequency such that $\omega/c \sim \beta_p$, so that the TM wave is dominant. Nevertheless, provided $\omega/c << \beta_p$ and in the limit $a/L_w \to 0$ the wire medium behaves effectively as a medium with extreme optical anisotropy and not as a medium with negative permittivity.

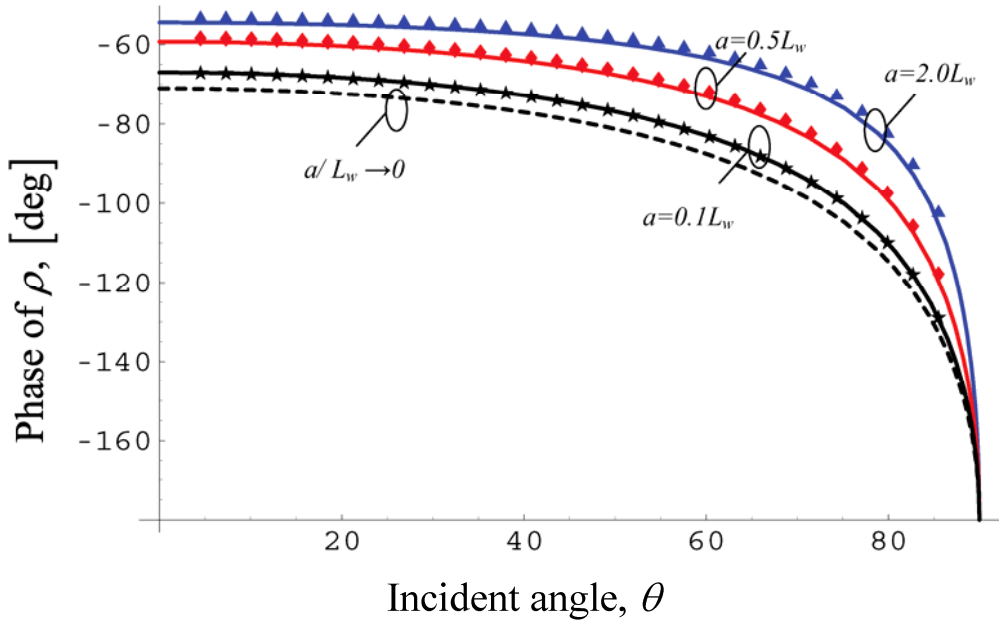

**Fig. 5** (Color online) Phase of the reflection coefficient as a function of the incident angle for different lattice constants. Solid lines: analytical model; Discrete symbols: exact result obtained with a full wave simulator. The substrate has permittivity $\varepsilon_h = 4.0$ and thickness $T$ such that $T\sqrt{\varepsilon_h}\omega/c = \pi/4$. The radius of the wires is $r_w = 0.05a$ and the tilt angle is $\alpha = 45^o$.

In the case in which the wires are densely packed, it is possible to obtain a simple expression for the reflection coefficient. From the previous discussion, it is expected that in such conditions the effect of the TM-mode is negligible. Thus, $\rho$ can be obtained by setting

$B_{TM}^{\pm} = 0$ in (15), (17) and by enforcing the classical boundary conditions. In this way, it is straightforward to verify that:

$$\rho \approx -\frac{j\left(e^{jk_{z,TEM}^{-}T} - e^{jk_{z,TEM}^{+}T}\right)\frac{\omega}{c}\sqrt{\varepsilon_h}\cos\alpha + \varepsilon_h\gamma_0\left(e^{jk_{z,TEM}^{+}T} + e^{jk_{z,TEM}^{-}T}\right)}{j\left(e^{jk_{z,TEM}^{-}T} - e^{jk_{z,TEM}^{+}T}\right)\frac{\omega}{c}\sqrt{\varepsilon_h}\cos\alpha - \varepsilon_h\gamma_0\left(e^{jk_{z,TEM}^{+}T} + e^{jk_{z,TEM}^{-}T}\right)}, \quad a/L_w \to 0 \quad (21)$$

It is stressed that the above expression for $\rho$ is coincident with that corresponding to a continuous material substrate characterized by the permittivity function (20).

To illustrate the described properties, we computed the reflection coefficient as function of the incident angle $\theta$, for $\alpha = 45°$, $\varepsilon_h = 4.0$, $T\sqrt{\varepsilon_h}\omega/c = \pi/4$ and for different values of the lattice spacing $a$. The ratio between the radius of the wires and the lattice constant is kept constant, $r_w = 0.05a$, so that the metal volume fraction is invariant. The simulation results are depicted in Fig. 5, showing again an excellent agreement between our analytical model and the results obtained with CST Microwave Studio$^{TM}$. It is seen that the response of the artificial substrate depends appreciably on the spacing between the wires. For large $a/L_w$ the spatial dispersion effects cannot be ignored, but when the wires are closely packed (e.g. for $a/L_w = 0.1$) the response of the material mimics very closely that of an ideal substrate following the permittivity model (20) (dashed line in Fig. 5), consistently with the previous discussion. It is relevant to mention that despite the structure not being invariant to a reflection relative to a plane normal to the *x*-direction, the phase characteristic of $\rho$ is an even function of the incidence angle $\theta$. This property is valid provided losses are negligible and all the diffraction modes are evanescent except for the fundamental spatial harmonic.

The propagation of electromagnetic waves in media with extreme optical anisotropy is very peculiar. In these materials, also known as waveguiding media [3], the energy tends to flow along the directions $\hat{\mathbf{x}}$ for which the dielectric function is infinite $\overline{\overline{\varepsilon}}_{eff}\cdot\hat{\mathbf{x}} = \infty$. In the present problem, this property can be easily understood by noting that the metallic wires can be regarded as subwavelength waveguides that sample the electromagnetic fields at the interface with air "pixel by pixel" [5]. In particular, since the coupling between adjacent wires is small because the wave is TEM, it is not surprising that the ratio between the electric and magnetic fields at the interface $z=0$ is independent of the angle of incidence, and only depends on the parameters characteristic of the medium with extreme anisotropy. In fact, it can be confirmed that for a substrate characterized by the permittivity function (20), one has:

$$Z_s \equiv -\frac{E_y}{H_x}\bigg|_{z=0} = j\eta_0 \frac{\cos\alpha}{\sqrt{\varepsilon_h}} \frac{j\left(e^{jk_{z,TEM}^- T} - e^{jk_{z,TEM}^+ T}\right)}{e^{jk_{z,TEM}^- T} + e^{jk_{z,TEM}^+ T}}$$
$$= j\eta_0 \frac{\cos\alpha}{\sqrt{\varepsilon_h}} \tan\left(\frac{\omega}{c}\sqrt{\varepsilon_h}T\sec\alpha\right), \qquad a/L_w \to 0 \qquad (22)$$

Notice that the above expression is independent of the angle of incidence, or more generally is independent of the transverse wave vector $k_x$. Thus, when the metallic wires are densely packed, the ratio between the tangential macroscopic electromagnetic fields at the interface with air is independent of the specific spatial variation of the incoming wave. In these conditions, the microstructured substrate is equivalent to an ideal "surface impedance" [26]. Such property is a generalization of the theory described in [24, 25] for the case of vertical wires ($\alpha = 0$).

It is well-known that an inductive surface impedance ($\text{Im}\{Z_s\} > 0$) supports guided modes that propagate along the surface [26]. Equation (22) shows that for long wavelengths the equivalent surface impedance has an inductive character. When $\frac{\omega}{c}\sqrt{\varepsilon_h}T\sec\alpha$ is slightly larger than $\pi/2$, the impedance becomes capacitive, and propagation is not possible. This suggests that the considered substrate supports TM-guided modes, whose stopband is defined by the condition $\frac{\omega}{c}\sqrt{\varepsilon_h}T\sec\alpha = \frac{\pi}{2}$, or equivalently the stopband begins when the thickness of the substrate $T$ is as large as $\frac{\lambda_0}{4\sqrt{\varepsilon_h}}\cos\alpha$ (the corresponding length of the wires $L_w$ is

$\frac{\lambda_0}{4\sqrt{\varepsilon_h}}$ ). Slightly below the band-gap lower frequency the guided modes are closely confined to the surface [26]. Thus, by tilting the metallic wires it seems possible to obtain a substrate that supports guided modes with subwavelength sizes.

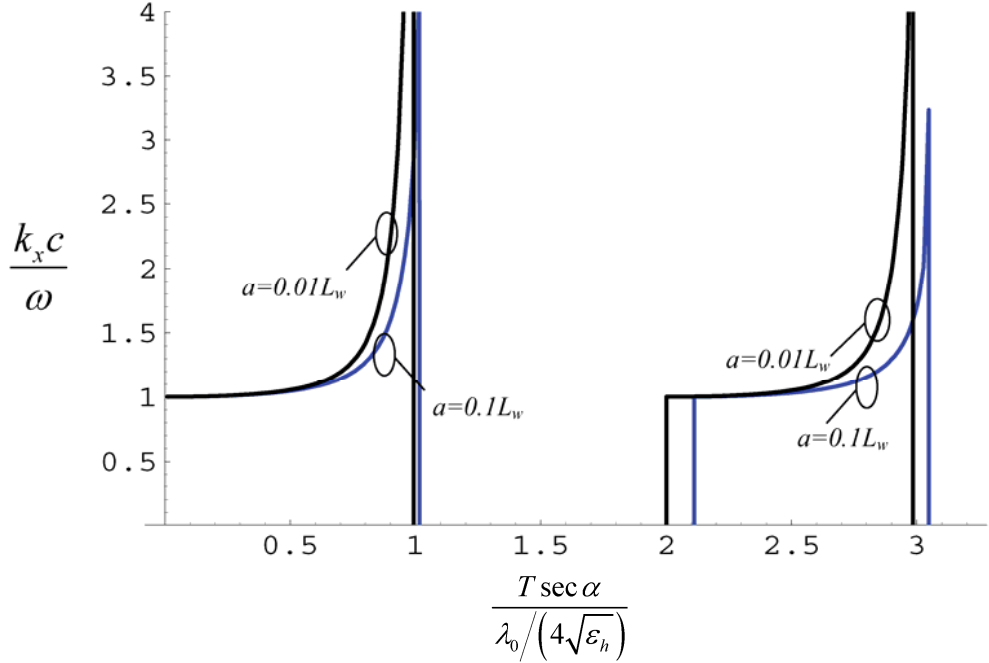

**Fig. 6** (Color online) Normalized propagation constant $k_x$ of the TM-surface wave modes (for propagation along the x-direction) as a function of the normalized thickness $T$ of the substrate and for a fixed frequency. The permittivity of the substrate is $\varepsilon_h = 2.2$, the radius of the wires is $r_w = 0.05a$, the tilt angle is $\alpha = 60°$, and the length of the wires is $L_w = T \sec \alpha$.

In order to confirm this hypothesis, we calculated the dispersion characteristic of the guided modes for a substrate formed by tilted wires with $\alpha = 60°$ and $\varepsilon_h = 2.2$, taking into account spatial dispersion effects. The dispersion characteristic is obtained by setting the determinant of the 5×5 linear system associated with the reflection problem equal to zero. Solving the corresponding equation numerically, we obtain $k_x = k_x(\omega)$, where $k_x$ is the propagation constant of the guided mode along the interface. The computed results are depicted in Fig. 6 for different values of the lattice constant $a$. It is seen that the guided modes become more attached to the textured surface (i.e. $k_x(\omega)$ increases) when the lattice constant

$a$ is decreased. Consistently with the surface impedance model (22), it is seen that the stopband emerges when the thickness of the substrate $T$ is close to $\frac{\lambda_0}{4\sqrt{\varepsilon_h}}\cos\alpha$, even if the wires are not closely packed. For $\alpha = 60º$ this implies that at a fixed frequency the thickness of the substrate can be reduced by a factor of $\cos 60º = 0.5$, as compared to the case of an unloaded substrate. Thus, by loading a substrate with tilted metallic wires it may be possible to reduce the size of the guided modes tail above the interface, and obtain a more compact dielectric waveguide.

## V. CONCLUSION

In this work, we derived new additional boundary condition to characterize the reflection of waves by an interface between a wire medium and a PEC ground plane. The new ABC is a consequence of the density of charge over the metallic wires being zero near the PEC interface. In addition, we generalized the results of [18] and proved that the ABC derived in [18] for a wire medium/air interface still applies when the wires are tilted with respect to the interface.

To illustrate the application of the new ABCs, we homogenized a substrate formed by an array of tilted metallic wires connected to a ground plane. Full wave simulations confirm the accuracy of our theories and demonstrate the role of the spatial dispersion effects. In particular, it was proven that spatial dispersion effects can be ignored only if the wires are very closely packed so that $a/L_w \approx 0$. In such circumstances, the wire medium behaves as a material with extreme optical anisotropy, and the microstructured substrate becomes equivalent to a "surface impedance". The effect of metallic losses can be easily incorporated into the proposed model by using the results of [10]. In particular, it can be verified that provided the skin depth of the metal is much smaller than the radius of the wires Ohmic losses are negligible. The results reported in this work besides of being of obvious theoretical interest, enable the modeling of the wire media connected to metallic surfaces. To conclude, we speculate that the studied microstructured surface may possibly be fabricated as an array of aligned metallic carbon nanotubes grown using the chemical vapor deposition method [27].

## APPENDIX A

Here we prove that for the case of a wire medium the TA-fields (1) can be characterized using the bulk medium homogenization model (11). As discussed in [23], in general such property is not valid. Our analysis is a generalization of the results derived in section VI.D of Ref. [23], which were obtained under the assumption that the wires are normal to the interface.

Consider the geometry depicted in Fig. 2, where the wire medium slab is illuminated by a plane wave. The microscopic fields $(\mathbf{e}, \mathbf{b})$ have the Floquet property along the directions parallel to the interface, i.e. $(\mathbf{e}, \mathbf{b}) e^{j \mathbf{k}_{\|} \cdot \mathbf{r}}$ is a periodic function of $x$ and $y$ with $\mathbf{k}_{\|} = (k_x, k_y, 0)$. It is well known that the inside the wire medium slab the electric field can be written as a superimposition of electromagnetic modes $\mathbf{e}_n$ ($n=1,2,\ldots$) supported by the unbounded crystal,

$$\mathbf{e}(\mathbf{r}) = \sum_n c_n \mathbf{e}_n (\mathbf{r}; \mathbf{k}_{\|}, k_{z,n}) \tag{A1}$$

where $c_n$ are some constants that define the amplitude of the modes. The electromagnetic mode $\mathbf{e}_n$ is associated with the wave vector $\mathbf{k}_n = \mathbf{k}_{\|} + k_{z,n} \hat{\mathbf{u}}_z$. Notice that the transverse component $\mathbf{k}_{\|}$ of the wave vector is invariant (independent of the mode $\mathbf{e}_n$) due to the periodicity of the structure along directions parallel to the interface [28].

Equation (A1) implies that inside the wire medium the TA-electric field is given by,

$$\mathbf{E}_{\text{av,T}}(z) = \sum_n c_n \mathbf{E}_{\text{av,T}}^{(n)}(z) \tag{A2}$$

where $\mathbf{E}_{av,T}^{(n)}(z;k_{z,n})$ is defined consistently with equation (1):

$$\mathbf{E}_{av,T}^{(n)}(z) = \frac{\cos\alpha}{a^2} \int_{-a/(2\cos\alpha)}^{a/(2\cos\alpha)} \int_{-a/2}^{a/2} \mathbf{e}_n(\mathbf{r}) e^{j\mathbf{k}_\parallel \cdot \mathbf{r}} \, dxdy \quad (A3)$$

To simplify the above formula, it is noted that $\mathbf{e}_n(\mathbf{r})e^{j\mathbf{k}_n \cdot \mathbf{r}}$ is independent of the specific value of $\mathbf{r} \cdot \hat{\mathbf{u}}_\alpha$ [9, 10]. Such property is a consequence of the invariance of the unbounded wire medium to arbitrary translations along the direction $\hat{\mathbf{u}}_\alpha$. Thus, we have that,

$$\begin{aligned}\mathbf{E}_{av,T}^{(n)}(z) &= e^{-jk_{z,n}z} \frac{\cos\alpha}{a^2} \int_{-a/(2\cos\alpha)}^{a/(2\cos\alpha)} \int_{-a/2}^{a/2} \mathbf{e}_n(\mathbf{r}) e^{j\mathbf{k}_n \cdot \mathbf{r}} \, dxdy \\ &= e^{-jk_{z,n}z} \frac{\cos\alpha}{a^2} \int_{-a/(2\cos\alpha)}^{a/(2\cos\alpha)} \int_{-a/2}^{a/2} \left(\mathbf{e}_n(\mathbf{r}) e^{j\mathbf{k}_n \cdot \mathbf{r}}\right)\Big|_{z=0} dxdy \end{aligned} \quad (A4)$$

where the second identity is a consequence of $\mathbf{e}_n(\mathbf{r})e^{j\mathbf{k}_n \cdot \mathbf{r}}$ being a periodic function of $x$ and being independent of the specific value of $\mathbf{r} \cdot \hat{\mathbf{u}}_\alpha$. Thus, the above result demonstrates that the dependence of $\mathbf{E}_{av,T}^{(n)}(z)$ on $z$ is simply of the form $e^{-jk_{z,n}z}$.

As mentioned before, the objective is to prove that in the long wavelength limit the vectors $\mathbf{E}_{av,T}^{(n)}$ can be characterized using the homogenization model [9, 10] of the bulk (unbounded) wire medium. In [10], the macroscopic *bulk* electric field associated with the electromagnetic mode $\mathbf{e}_n$ was defined by,

$$\mathbf{E}_{av}^{(n)} = \frac{1}{V_{cell}} \int_\Omega \mathbf{e}_n(\mathbf{r}) e^{j\mathbf{k}_n \cdot \mathbf{r}} d^3\mathbf{r} \quad (A5)$$

where $\Omega$ is the unit cell of the periodic crystal and $V_{cell}$ is the corresponding volume. Since the wire medium can be regarded as a simple cubic lattice of wires with lattice constant $a$, we have that $V_{cell} = a^3$.

In order to relate $\mathbf{E}_{av,T}^{(n)}(z)$ with $\mathbf{E}_{av}^{(n)}$, we note that the unit cell of the unbounded medium can be taken equal to $\Omega = \{(x,y,0) + s\hat{\mathbf{u}}_\alpha : |x| \leq a/(2\cos\alpha), |y| \leq a/2, |s| \leq a/2\}$ where $\hat{\mathbf{u}}_\alpha$ is as in Fig. 2. Since the element of volume is $d^3\mathbf{r} = \cos\alpha \, dxdyds$, eq. (A5) is equivalent to:

$$\mathbf{E}_{av}^{(n)} = \frac{\cos\alpha}{a^3} \int_{-a/2}^{a/2} \left( \int_{-a/(2\cos\alpha)}^{a/(2\cos\alpha)} \int_{-a/2}^{a/2} \mathbf{e}_n(\mathbf{r}) e^{j\mathbf{k}_n \cdot \mathbf{r}} dxdy \right) ds \tag{A6}$$

But from the second identity of (A4) it follows that:

$$\mathbf{E}_{av}^{(n)} = \frac{\cos\alpha}{a^3} \int_{-a/2}^{a/2} \left( \int_{-a/(2\cos\alpha)}^{a/(2\cos\alpha)} \int_{-a/2}^{a/2} \left( \mathbf{e}_n(\mathbf{r}) e^{j\mathbf{k}_n \cdot \mathbf{r}} \right)\bigg|_{z=0} dxdy \right) ds \tag{A7}$$

It is simple to verify that the integrand in the above equation is independent of the variable $s$. Hence, using again equation (A4), we finally find that:

$$\mathbf{E}_{av,T}^{(n)}(z) = \mathbf{E}_{av}^{(n)} e^{-jk_{z,n}z} \tag{A8}$$

Thus, the transverse averaged fields associated with an electromagnetic mode are completely characterized by the corresponding bulk medium fields. In particular, from (A2) the TA-electric field inside the wire medium can be written as,

$$\mathbf{E}_{av,T}(z) = \sum_n c_n \mathbf{E}_{av}^{(n)} e^{-jk_{z,n}z} \tag{A9}$$

This result is mathematically exact and only assumes that the wires are not parallel to the interface. In the long wavelength limit, the above series is truncated and only the fundamental modes that propagate at low frequencies are retained. The macroscopic field $\mathbf{E}_{av}^{(n)}$ associated with these modes is characterized by the dielectric function (11). Hence, equation (A9) demonstrates that in the long wavelength limit the TA-fields inside the wire medium slab are completely characterized by the dielectric function (11), as we wanted to establish.

## APPENDIX B

Here, we analyze with more detail the boundary conditions at the PEC interface when the wires are normal to the interface, $\alpha = 0$. From (6), it is known that in such conditions the ABC is equivalent to $\hat{\mathbf{u}}_z \cdot \frac{d\mathbf{J}_{d,av}}{dz} = 0$. For a scattering problem, the total macroscopic electric field inside the wire medium can be written as a superimposition of TE, TM and TEM modes [9],

$$\mathbf{E}(\mathbf{r}) = \sum_{n=TM,TE,TEM} c_n^+ \mathbf{E}_n(\mathbf{r};\mathbf{k}_\parallel,k_z) + c_n^- \mathbf{E}_n(\mathbf{r};\mathbf{k}_\parallel,-k_z), \qquad \alpha=0 \qquad (B1)$$

where $c_n^\pm$ are unknown coefficients. For simplicity, it is assumed that the PEC plane is placed at $z=0$. We claim that when $\alpha=0$, the condition $\hat{\mathbf{u}}_z \times \mathbf{E} = 0$ and the ABC $\hat{\mathbf{u}}_z \cdot \dfrac{d\mathbf{J}_{d,\mathrm{av}}}{dz} = 0$ imply that $c_n^- = -c_n^+$. Indeed, because of invariance of the wire medium under the transformation $R_z:(x,y,z)\to(x,y,-z)$ the averaged electromagnetic modes satisfy $\mathbf{E}_n(\mathbf{r};\mathbf{k}_\parallel,-k_z) = R_z \cdot \mathbf{E}_n(R_z \cdot \mathbf{r};\mathbf{k}_\parallel,k_z)$. Thus, (B1) is equivalent to:

$$\mathbf{E}(\mathbf{r}) = \sum_{n=TM,TE,TEM} c_n^+ \mathbf{E}_n(\mathbf{r};\mathbf{k}_\parallel,k_z) + c_n^- R_z \cdot \mathbf{E}_n(R_z \cdot \mathbf{r};\mathbf{k}_\parallel,k_z), \qquad \alpha=0 \qquad (B2)$$

Hence, it is clear that if $c_n^- = -c_n^+$ the boundary condition $\hat{\mathbf{u}}_z \times \mathbf{E} = 0$ is satisfied at $z=0$. On the other hand, because of similar symmetry arguments, the averaged microscopic current (defined as in equation (3)) associated with each mode also satisfies $\mathbf{J}_{d,\mathrm{av},n}(\mathbf{r};\mathbf{k}_\parallel,-k_z) = R_z \cdot \mathbf{J}_{d,\mathrm{av},n}(R_z \cdot \mathbf{r};\mathbf{k}_\parallel,k_z)$. Therefore, the total averaged microscopic current is given by:

$$\mathbf{J}_{d,\mathrm{av}}(\mathbf{r}) = \sum_{n=TM,TE,TEM} c_n^+ \mathbf{J}_{d,\mathrm{av},n}(\mathbf{r};\mathbf{k}_\parallel,k_z) + c_n^- R_z \cdot \mathbf{J}_{d,\mathrm{av},n}(R_z \cdot \mathbf{r};\mathbf{k}_\parallel,k_z), \qquad \alpha=0 \qquad (B3)$$

But since $\hat{\mathbf{u}}_z \cdot \mathbf{J}_{d,\mathrm{av},n}(\mathbf{r};\mathbf{k}_\parallel,k_z) + \hat{\mathbf{u}}_z \cdot \mathbf{J}_{d,\mathrm{av},n}(R_z \cdot \mathbf{r};\mathbf{k}_\parallel,k_z)$ is an even function of $z$, it is clear that when $c_n^- = -c_n^+$ the ABC $\hat{\mathbf{u}}_z \cdot \dfrac{d\mathbf{J}_{d,\mathrm{av}}}{dz} = 0$ is verified at $z=0$. This demonstrates our claim. In particular, the previous analysis shows that when the wires are normal to the ground plane the different electromagnetic modes verify *independently* the PEC boundary condition [25],

$$\hat{\mathbf{u}}_z \times \mathbf{E}_n = 0, \quad n=\text{TE, TM, TEM}, \qquad \alpha=0 \qquad (B4)$$

Thus, when $\alpha=0$ the new ABC and the classical boundary condition $\hat{\mathbf{u}}_z \times \mathbf{E} = 0$, are completely equivalent to the set of boundary conditions defined by (B4).

ACKNOWLEDGEMENT

This work was funded by Fundação para Ciência e a Tecnologia under project POSC/EEACPS/61887/2004.